# Super-efficiency of Listed Banks in China and Determinants Analysis (2006-2021)

Yun Liao [1]    Ruihui Xu[2*]

[1] *Author affiliation: Credit Management School, Guangdong University of Finance, Guangzhou, China. Email: sihtyh@foxmail.com*

[2] *Author affiliation: Research Institute, The People's Bank of China, No. 32 Chengfang Street. Xicheng District, Beijing, China*

\* Corresponding Email:  xruihui@pbc.gov.cn

# Super-efficiency of Listed Banks in China and Determinants Analysis (2006 to 2021)

Abstract: This study employs the annual unbalanced panel data of 42 listed banks in China from 2006 to 2021, adopts the non-radial and non-oriented super-efficiency Data envelopment analysis (Super-SBM-UND-VRS based DEA) model considering NPL as undesired output. Our results show that the profitability super-efficiency of State-owned banks and Rural/City Commercial Banks is better than that of Joint-stock Banks. In terms of intermediary efficiency(deposit and loan), state-owned banks have advantage on other two type of banks. The determinants analysis shows that all type of banks significantly benefits from the decrease of ownership concentration which support reformation and IPO. Regional commercial banks significantly benefit from the decrease of customer concentration and the increase of reserves. On the other hand, State-owned banks should increase its loan to deposit ratio while joint-stock banks should do the opposite.

# 1.Introduction

China's banking industry is in a period of deepening reform. With the lifting of the 30% lower limit on loan interest rates by the central bank in 2013, China's banking industry has entered a new stage of interest rate marketization. On the other hand, the loan to deposit ratio (LDR) ban was lifted in 2015, become a liquidity risk monitoring indicator. The difference between the benchmark one-year loan interest rate and the one-year treasury bond rate has decreased by about 200bps from 2006 to 2021, total asset of banking sector in china had increased from 39 trillion to 379 trillion RMB Yuan in this period.

How to accurately evaluate the efficiency of commercial banks has always been an important topic faced by regulators, bank managers, institutional and individual investors. There are many methods to study the efficiency of commercial banks, the representative method of parametric method is stochastic frontier analysis (SFA), which requires pre-set functional forms and is more inclined to study central tendencies. The representative of non-parametric method is data envelopment analysis (DEA), that is, only the linear relationship between input and output is pre-set, and the efficiency is allowed to change with time. Scholars examined China's banking sector and get some distinct results, Berger, Hasan et al. (2009) point out minority foreign ownership of the Big Four will likely improve performance significantly, Jiang, Yao et al. (2013) shows that public traded banks have better performance, Wang, Huang et al. (2014) use two-stage DEA method and find out the difference of efficiency between SOBs and JSBs are decreasing after 2003, Zhang, Cai et al. (2016) suggest that an increase in the NPLs ratio raises riskier lending. However, there are a lot of changes in recent years' reform in china banking sector, it is necessary to reinvestigate how these reformations shaped china banking sector.

Traditional DEA models such as CCR and BBC suffer from some limitations such as the difficulty in statistical testing the results; Failure to clearly identify ways to improve efficiency; Pure efficiency does not accurately reflect the risk-taking of banks. This paper adopts data envelopment analysis including non-directional and non-radial super efficiency based on relaxation conditions. The Super-SBM-UND-VRS Based DEA model provides an efficiency score greater than one for the unit, which is determined by the slack conditions of input and output, and can clearly indicate the route to improve efficiency. However, super-efficiency data envelopment analysis may have infeasible issues, especially when the size of some companies is extremely small or extremely large relative to the sample. The data envelopment analysis model based on slack conditions can provide the advantage of discrimination power and detect more sources of inefficiency, the disadvantage is that the study may fails to identify and consider internal sub-production processes. To solve this problem, we employ profit approach and Intermediary approach to evaluate the efficiency, profit approach demonstrate the value creation in shareholders' view, intermediary approach measures save and loan efficiency considering not performing loan(NPL) as undesired output. These efficiencies provide a clear inside view into the three types of China's listed banks. Our paper maybe the first and newest paper to investigate the two efficiencies in china banking sector and their determinants. Our study may shed light on the performance analysis in the field.

In the second stage, we use decomposition of MI index, catch-up effect and frontier effect, to evaluate the development trend and competitive positioning of individual commercial banks. In addition, further analyse of determinants reveal some useful information about China listed bank efficiency, give a big picture on China's banking industry reform.

Section 2 reviews literature on China's bank efficiency and determinants analysis in China banking sector. Section 3 outlines our empirical methodology: SUPER-SBM-DEA-VRS (non-radial, non-directional) model and data on the China's listed banks. Section 4 displays empirical results about individual banking efficiency and give comprehensively analyse. In section 5 the empirical analysis carries out by panel regression analysis on the catch-up effect. Section 6 provide some implications and suggestions.

**2. Literature Review**

The value of Data Envelope Analysis lies in the ability to relatively assess individual efficiency or the efficiency of Decision-Making units within goals. This method was first proposed by Charnes, Cooper et al. (1978). This model assumes constant returns to scale and is widely used in the industrial field. Sherman and Gold (1985) is the first paper to apply DEA to study bank efficiency, the classic CCR model adopt the acronym of the three economists is used to compare operational efficiency among the 14 branches of a savings bank. Financial institution efficiency studies various ways of defining inputs and outputs in financial services. For example, the intermediation method defines loans and deposits as outputs, while other liabilities, labor, and physical capital are considered inputs.

Grigorian and Manole (2006) used data Envelopment Analysis (DEA) to analyze bank efficiency in their study on the ownership structure of banks in Eastern European countries during the transition period from 1995 to 1998, and used the Tobit regression model with constraints to draw conclusions: Banks with low asset-liability ratios are more efficient, large banks are more efficient than small and medium-sized banks, and foreign ownership and restructuring can significantly improve efficiency. Berger, Hasan et al.

(2009) pointed out through the study of Chinese banks from 1994 to 2003 that minority foreign ownership may significantly improve the efficiency of the four state-owned banks.

Chinese banking efficiency studies have used different methods (parametric and non-parametric) to estimate different efficiency indicators (technical efficiency, cost efficiency and profit efficiency). The consensus of scholars is that the reform of more than two decades has improved the performance of Chinese banks. Some studies believe that joint-stock commercial banks are more efficient than state-owned commercial banks. For example, Zhang (2003)，Yao, Feng et al. (2004), Wang, Huang et al. (2014), Zhu, Wang et al. (2016) . On the other hand, some studies have pointed out that state-owned commercial banks are more efficient, such as Yang and Zhang (2007); Gan (2007).

There are also a large number studies on the determinants of banking efficiency , such as loan concentration, non-interest income and ownership concentration effect on the banking efficiency and risk. Zhao and Shen (2016), Li (2021) believed that non-interest income(Nii) had a positive effect on the overall risk of banks; Zhao (2021) believed that Nii had a negative effect; Jiang and Li (2019), Gao, Wang et al. (2021)believed that there may be a threshold effect about Nii. Wang, Wang et al. (2020) believed that the loan-to-deposit ratio has an inverted U-shaped relationship with the profitability of commercial banks, and Li and Wang (2018) believed that the cancellation of the loan-to-deposit ratio constraint may have a positive effect on the efficiency improvement of banks. The results of this paper can be used to verify relevant theories.

The research in the field has achieved some important results, but there are some unsolved problems. First, in the analysis, single risk indicators such as Z value, non-performing loan ratio, MES, etc., or interest margin, ROE, etc., are often used, which makes it difficult to consider profit and risk together. Second, the traditional DEA model is adopted, which makes it impossible to compare the units on the production boundary with each other, it is also impossible to conduct further analysis on its determinants; Third, the decomposed results of MI index are difficult to interpret. For example, technical efficiency and scale efficiency are difficult to be explained by the relevant characteristics of individual banks; In view of this, this paper uses the super-efficiency SBM-V model to obtain comparability and analysability.

By using the super-efficiency SBM model, two important attributes of SBM can be obtained: It can not only measure the slack conditions of input and output, but also compare the units which are located at the boundary of the production feasible set, so the model is more accurate and effective (compared with other DEA models). Further analysis was conducted by using these generated indicators.

## 3.Research methods and Data Sources

### *3.1 Super-efficiency Super-SBM-V model based on undesired output*

The original DEA model (CCR model) was developed under the CRS (Constant Return to Scale) assumption, which means that "a t-fold increase in inputs will lead to a t-fold increase in output". Banker, Charnes et al. (1984) developed the BCC model using the VRS assumption. The Variable Return to Scale (VRS) assumption implies that "an equal proportional increase in factor inputs produces a greater (or smaller) equal proportional increase in output". The Slack Based Measure DEA of Tone (2001) incorporates the slack improvement of inefficiency to ensure that the final efficiency measurement is effective, and meets the monotonicity of unit invariance with output-slack and input-slack conditions. This model solves the defects of input and output directionality (CCR and BCC models) in efficiency measurement, and develops the non-radial and non-directional SBM-VRS model. The super efficiency model proposed by

Andersen and Petersen (1993) is mainly to solve the problem that it is difficult to compare the efficiency of two units located on the boundary of the feasible set of production.

The undesirable output model proposed by Färe, Grosskopf et al. (1994) further solves the problems in efficiency measurement such as environmental pollution by-products. Fujii, Managi et al. (2018) Fujii, Managi et al. (2018) conducted DEA analysis on the efficiency of commercial banks in 28 EU countries from 2005 to 2014, using per capita expenses, fixed assets and deposits as inputs and loans, other profitable assets and non-interest income as outputs in model analysis. The above studies all use the SBM model with undesired output, but not use the super efficiency model.

Due to the complex differences in the regulatory environment, operating environment and the difficulty of obtaining funds faced by banks of different sizes, for the sake of fairness, the data envelopment analysis of this paper uses variable returns to scale (VRS) to measure efficiency. In essence, commercial banks of similar sizes are comparable. However, its disadvantage is that the pooling factor regression using the efficiency calculated by this method may produce certain deviations, which needs to be paid attention to. Therefore, the panel factor regression in the second stage of this paper adopts difference panel regression to eliminate the scale bias.

Considering that commercial banks, as a for-profit organization, obtain net profits through a series of input capital and labour, intermediate products include deposits, loans, interest and non-interest income. Non-performing loans (NPL) of banks is an undesirable output. On the other hand, in order to analyse the capital use efficiency of banks, especially the efficiency of providing financial services for the real economy and residents, this paper refers to many literatures and chooses tier one capital, interest expense and non-interest expense as the input. In the production approach (PA), the profit efficiency of commercial banks is evaluated based on the perspective of risk and return.

Two variables are used in the output: the first output is the net profit returned to the parent after the consolidated statement; The second output is the undesirable by-product, non-performing loans. The intermediary method (hereinafter referred to as IA) considers deposits and loans, and mainly analyses the intermediary efficiency of banks.

The following is a prototype of the SBM model with an undesirable output：

[SBM-Undesirable][3]

$$\rho^* = \min \frac{1 - \frac{1}{m}\sum_{i=1}^{m} \frac{s_i^-}{x_{i0}}}{1 + \frac{1}{s_1+s_2}\left(\sum_{r=1}^{s_1} \frac{s_r^g}{y_{r0}} + \sum_{r=1}^{s_2} \frac{s_r^b}{y_{r0}}\right)}$$

$$S.T \begin{cases} x_0 = X\lambda + s^- \\ y_0^g = Y^g\lambda - s^g \\ y_0^b = Y^b\lambda + s^b \\ s^-, s^g, s^b, \lambda \geq 0 \end{cases} \quad (1)$$

Considering that there are m inputs (x), $s_1$ kind of expected outputs ($y^g$) and $s_2$ kinds of undesired outputs ($y^b$), output y is ($y^g$, $y^b$), which represents expected outputs and undesired outputs respectively. $s^- \epsilon R^m$ and $s^b \epsilon R^{s_2}$ represents excess of input and undesirable output, and $s^g \epsilon R^{s_1}$ as shortage of output. The current decision unit is efficient if and only $\rho = 1$, and $s^-$, $s^g$, $s^b$ are zeros. When the three relaxation conditions are not all zeros, the decision-making unit lacks efficiency. The form of the improved super efficiency SBM model is:

---

[3] Cooper, W. W., L. M. Seiford and K. Tone (2007). <u>Data Envelopment Analysis</u>, Springer Books., page318

$$\rho^* = \min \frac{\frac{1}{m}\sum_{i=1}^{m}\frac{\bar{x}}{x_{i0}}}{\frac{1}{s_1+s_2}\left(\sum_{r=1}^{s_1}\frac{\bar{y}_r^g}{y_{r0}} + \sum_{r=1}^{s_2}\frac{\bar{y}_r^b}{y_{r0}}\right)}$$

$$S.T\begin{cases} \bar{x} \geqslant \sum_{j=1,\neq 0}^{n}\lambda_j x_j \\ \bar{y}_r^b \geqslant \sum_{j=1,\neq 0}^{n}\lambda_j y_j^b \\ \bar{y}_r^g \leqslant \sum_{j=1,\neq 0}^{n}\lambda_j y_j^g \\ \bar{x} \geqslant x_0, \bar{y}_r^b \geqslant y_0^b \text{ and } \bar{y}_r^g \leqslant y_0^g \\ \sum_{j=1,\neq 0}^{n}\lambda_j = 1 \quad (VRS\ condition) \\ y \geqslant 0, \lambda \geqslant 0 \end{cases} \quad (2)$$

In order to solve the problem that the method of Super-SBM in some cases has no feasible solution, this paper refers to the method of Fang, Lee et al. (2013) for two-stage solution. The super-efficiency are defined as:

$$SE = \begin{cases} 1 - \frac{1}{m}\sum_{1}^{m}\frac{s_i^-}{x_i} - \frac{1}{k}\sum_{1}^{m}\frac{s_r^b}{y_i}; & if\ SE < 1 \\ 1 + \frac{1}{m}\sum_{1}^{m}\frac{s_i^-}{x_i} + \frac{1}{k}\sum_{1}^{m}\frac{s_r^b}{y_i}; & if\ SE > 1 \end{cases} \quad (3)$$

### *3.2 Malmquist index based on super-efficiency DEA*

Since the DEA model is difficult to compare DMU efficiency in different stages, Färe, Grosskopf et al. (1994) used the radial DEA method to calculate the Malmquist index of 17 OECD countries. The Malmquist index (MI) is decomposed into the technical Efficiency Change (Efficiency Change) and the production technology change (Technological Change) of DMU in two periods. Among them, the change in technical efficiency (EC) reflects the change in the production efficiency of the DMU or the catch-up effect, and TC reflects the degree of movement of the production frontier or the frontier-shift-effect. The main form of MI index is as follows:

$$\begin{cases} C = \dfrac{\delta^2\left((x_0,y_0)^2\right)}{\delta^1\left((x_0,y_0)^1\right)} \\ F = \left[\dfrac{\delta^1\left((x_0,y_0)^1\right)}{\delta^2\left((x_0,y_0)^1\right)} \times \dfrac{\delta^1\left((x_0,y_0)^2\right)}{\delta^2\left((x_0,y_0)^2\right)}\right]^{1/2} \\ MI = C * F = \left[\dfrac{\delta^1\left((x_0,y_0)^2\right)}{\delta^1\left((x_0,y_0)^1\right)} \times \dfrac{\delta^2\left((x_0,y_0)^2\right)}{\delta^2\left((x_0,y_0)^1\right)}\right]^{1/2} \end{cases} \quad (4)$$

Given that it represents the efficiency of the DMU in period t2 compared with each DMU in period t1, the Malmquist index essentially reflects the progress level of the DMU. Since the radial DEA model does not consider slack variables, this paper uses the non-radial, non-oriented SUPER-SBM-V model to calculate the Malmquist index of each bank.

*3.3 Data and Variable Selection*

We use the annual report data of 42 listed banks as the main data source. The sample includes the annual reports of banks from 2006 to 2021. According to the classification of Shanghai Stock Exchange, the sample includes 6 large state-owned banks, 9 joint-stock banks, 17 urban commercial banks and 10 rural commercial banks, with a total of 353 observed values. The full name and abbreviation of listed bank in China are provided in the appendix.

We use tier one capital as input other than fixed assets, and the final net profit of banks as output. Undesirable output is non-performing loans (NPL), which follows Classification of loans and debt securities that are in a delinquent state (90 days overdue) or have a very low probability of repayment. As shown in Table I, the production approach (PA) is measured by the net profit excluding non-recurring gains and losses attributable to the parent company as the final target product; The intermediary approach (IA) evaluates the operating efficiency of banks by taking deposits and loans as the intermediary products of banks, and take non-performing loans as undesirable products.

Table 1: Input and output used in the SUPER-SBM model

| Category | X1 | X2 | X3 | X4 | Y1 | Y2 | YB (非期望产出) |
|---|---|---|---|---|---|---|---|
| PA | Tier one capital | Interest expense | Non-interest expense | | Net profit | | NPL |
| IA | Tier one capital | Interest expense | Non-interest expense | | Deposit | Loan | NPL |
| PA1 | Total asset | Interest expense | Non-interest expense | | Net profit | | NPL |
| IA1 | Fixed asset | Interest expense | Non-interest expense | | Deposit | Loan | NPL |
| PA as production approach；IA as intermediary approach | | | | | | | |

## *3.4 Descriptive statistics of the major variables*

Table II shows the descriptive statistics of the main inputs and outputs in the super-efficiency analysis, and it can be seen that the 42 listed commercial banks basically cover large, medium and small-scale commercial banks. Appendix 1 shows some detail information about the listed bank in china.

Table 2: Description of Super-efficiency analysis Variables

| Variables(100M Yuan) | Represent for | Obs. | median | std | min | P25 | P50 | P75 | max |
|---|---|---|---|---|---|---|---|---|---|
| netprofit | net profit | 353 | 481.3 | 714.53 | 5.94 | 43.74 | 180.23 | 540.46 | 3458.99 |
| asset | total asset | 353 | 50055.3 | 69447.21 | 579.87 | 5583.86 | 20482.25 | 62896.06 | 351713.83 |
| ie | interest expenses | 353 | 858.49 | 1031.17 | 9.17 | 117.02 | 433.27 | 1223.94 | 4715.38 |
| oe | non-interest expenses | 353 | 759.23 | 1033.07 | 7.32 | 78.2 | 284.05 | 1070.42 | 5191.98 |
| coreasset | tie-one capital | 353 | 3340.33 | 5138.68 | 23.61 | 304.72 | 1163.78 | 4157.26 | 28863.8 |
| save | deposit | 353 | 35328.53 | 52178.21 | 438.58 | 3522.92 | 13032.2 | 37590.6 | 264418 |
| loan | loan | 353 | 26594.57 | 38953.95 | 255.04 | 2442.05 | 10247.3 | 32721.6 | 206672 |
| npl | non-performing loan | 353 | 385.47 | 588.79 | 1.31 | 31.37 | 101.24 | 522.17 | 2942.64 |

## 4. Empirical Study on the Super-efficiency of Banks

### *4.1 Full Time Series Analysis of Super efficiency of listed Banks (2006-2021)*

Figure 1 shows the PA efficiency and IA efficiency rankings of listed commercial banks from 2006 to 2021, and the appendix shows the risk-based super-efficiency trends of all listed banks from 2006 to 2021 calculated by the production method (PA) and the intermediary method (IA).

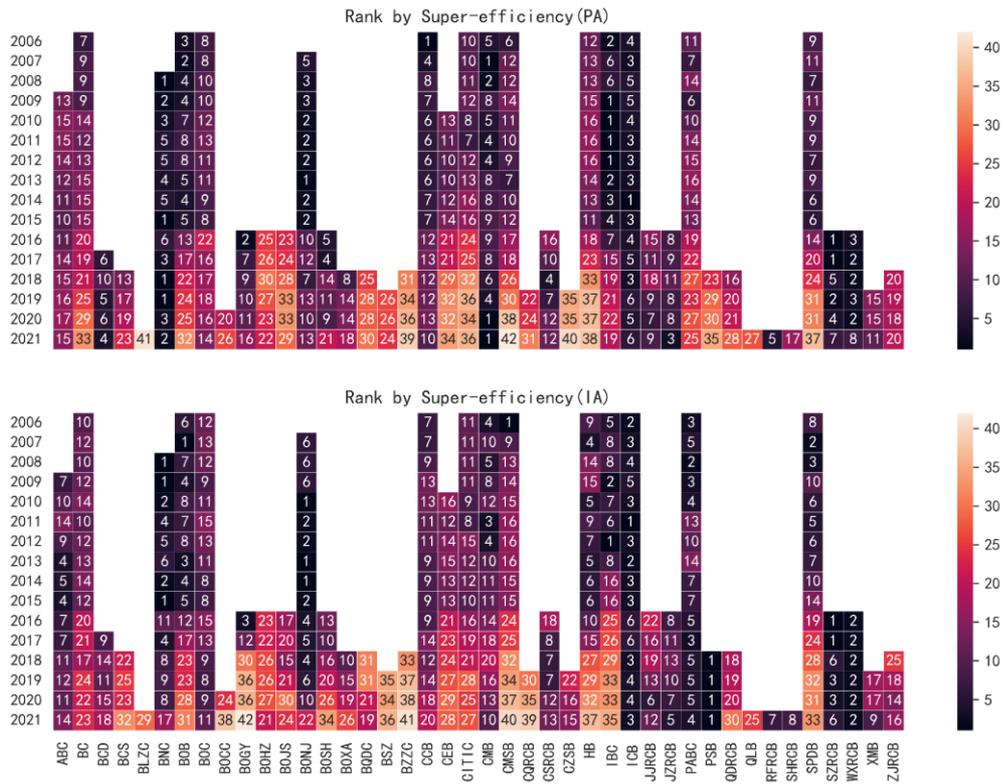

Figure 1: Rank by Super-efficiency (PA) Listed Bank (2006-2021)

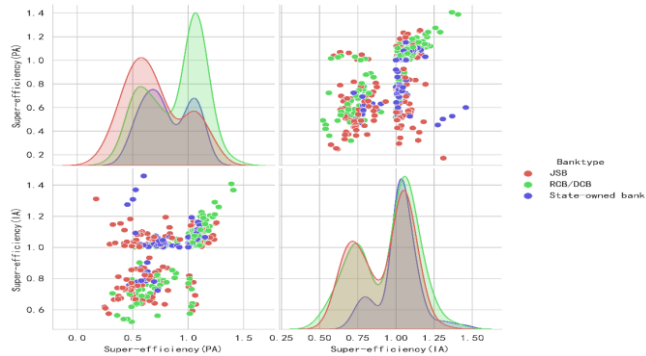

Figure 2: Distribution of Super-efficiency of Listed Bank (2006-2021)

From the super efficiency ranking diagram in Figure 1 and the distribution scatter diagram in Figure 2, it can be seen that the three types of banks all have the characteristics of bimodal distribution, that is, each type of banks has both excellent typical banks and relatively poor banks. From the sample, the banks in the first year of IPO usually rank high in super efficiency, which may be because the asset structure of the banks has been optimized after replenishing capital, and the non-performing loan ratio has also decreased.

Through the Wilcoxon-Mann-Whitney rank-sum statistics and t-test, we can have the ranking of the efficiency of three types of commercial banks in a statistical sense. The test results are shown in Table III.

Table 3: Comparative Statistics of Listed Banks by Type(2006-2021)

| Super-efficiency (PA) test | H0 | H1 | Rank-Sum-Stat | p-value |
|---|---|---|---|---|
| [1, 2] | 1>2 | 1<2 | 6596 | 0.9997 |
| [1, 3] | 1>3 | 1<3 | 3558 | 0.0804 |
| [2, 3] | 2>3 | 2<3 | 4322 | 0.0000 |
| | H0 | H1 | t-test | p-value |
| [1, 2] | 1<2 | 1>2 | 3.2565 | 0.0007 |
| [1, 3] | 1<3 | 1>3 | -1.4165 | 0.9208 |
| [2, 3] | 2<3 | 2>3 | -4.6307 | 1.0000 |
| Super-efficiency (IA) test | H0 | H1 | Rank-Sum-Stat | p-value |
| [1, 2] | 1<2 | 1>2 | 6564 | 0.0004 |
| [1, 3] | 1<3 | 1>3 | 5022 | 0.0028 |
| [2, 3] | 2<3 | 2>3 | 6230 | 0.5969 |
| | H0 | H1 | t-test | p-value |
| [1, 2] | 1>2 | 1<2 | 4.6004 | 1.0000 |
| [1, 3] | 1>3 | 1<3 | 4.0465 | 1.0000 |
| [2, 3] | 2>3 | 2<3 | -0.0289 | 0.4885 |

1、2、3 represented as State-owned Banks, JSBs and DCB/RCB respectively.

Two tests (t-test and rank-sum-stat) are used to find the super efficiency under the production method: state-owned large and regional commercial banks are better than joint-stock commercial banks with a confidence level of 99%. Under the comparison of intermediary method, the super efficiency of large state-owned commercial banks is better than that of regional commercial banks and joint-stock commercial banks with a confidence level of 99%, so it can be concluded that large state-owned commercial banks are generally better than the other two types of banks in the efficiency of financing intermediary. On the other hand, joint-stock commercial banks have a certain weakness in PA efficiency, a possible reason is that they could be disadvantage in national-wide competition with state-owned bank. On the other hand, they have certain homogeneity with rural or city commercial banks in regional competition. Except China Merchants Bank, other joint-stock commercial banks have mediocre performance.

## 4.2 Super efficiency decomposition -Inefficiency analysis of commercial banks

A significant feature of the Super-SBM-VRS model is the ability to analyse the inefficiency of each input variable separately. Table 4 shows the average super efficiency and inefficiency of listed banks. The average super efficiency of large state-owned banks, joint-stock banks and regional banks measured by PA (production method) is 0.81, 0.70 and 0.89 respectively. The average super-efficiency of large state-owned banks, joint-stock banks and regional banks measured by IA (intermediary method) is 1.02, 0.92 and 0.94, respectively.

The PA efficiency of state-owned and regional banks is significantly better than that of joint-stock banks, and the IA efficiency of large state-owned banks is significantly better than that of joint-stock and regional banks. In terms of inefficiency PA, there is a certain gap between joint-stock banks and other two types of banks in terms of interest expenses, operating expenses and non-performing loans. In IA inefficiency, large state-owned banks are significantly better than the other two types of banks in terms of inefficiency except loan inefficiency.

Table 4: average super efficiency and inefficiency of listed banks(2006-2021)

| PA | SOB | JSB | RCB/CCB |
|---|---|---|---|
| DmuYear | 81 | 127 | 145 |
| PA | 0.8135 | 0.7004 | 0.8874 |
| coreasset_slack | 0.0442 | 0.1123 | 0.0483 |
| ie_slack | 0.1609 | 0.2722 | 0.1422 |
| oe_slack | 0.1241 | 0.2016 | 0.1365 |
| netprofit_slack | 0.04 | 0.045 | 0.0278 |
| npl_slack | 0.1618 | 0.2385 | 0.1471 |
| IA | SOB | JSB | RCB/CCB |
| Dmuyear | 81 | 127 | 145 |
| IA | 1.0241 | 0.9179 | 0.9441 |
| coreasset_slack | 0.0259 | 0.0544 | 0.0793 |
| ie_slack | 0.0407 | 0.1096 | 0.1242 |
| oe_slack | 0.0076 | 0.0529 | 0.0816 |
| save_slack | 0.1129 | 0.1134 | 0.0708 |
| loan_slack | 0.0648 | 0.0459 | 0.0253 |
| npl_slack | 0.0237 | 0.0908 | 0.1011 |

Figure 3 is the PA analysis of listed banks. It can be seen from the figure that the main sources of inefficiency of joint-stock banks are interest expense and NPL. For example, the inefficiency of interest expenditure of Minsheng Bank, Huaxia Bank, China Citic Bank, Everbright Bank and Shanghai Pudong Development Bank are all close to or even over 0.5. Similarly, the NPL inefficiency of Minsheng Bank, Huaxia Bank, China Citic Bank and Shanghai Pudong Development Bank exceeds or approaches 0.5.

Among large state-owned banks, Postal Savings Bank has the highest operating expense inefficiency, reaching 0.43. Among regional commercial banks, Bank of Zhengzhou, Bank of Lanzhou, Bank of Qingdao, Bank of Hangzhou, Bank of Changsha and Bank of Qilu have high operating expense inefficiency. Among large state-owned banks, Bank of Communications has the highest inefficiency of NPL, reaching 0.47. Among regional commercial banks, Bank of Zhengzhou, Bank of Lanzhou, Qingdao rural Commercial Bank and Bank of Beijing have high inefficiency of NPL, which is worthy of vigilance.

Figure 4 shows the probability density of inefficiency (PA) of listed banks. It can be seen from the figure that the inefficiency of regional commercial banks is between that of large state-owned banks and joint-stock banks, which also explains the overall superior super efficiency (PA) of these two types of banks to joint-stock banks.

Figure 3: PA analysis chart of listed banks (2006-2021)

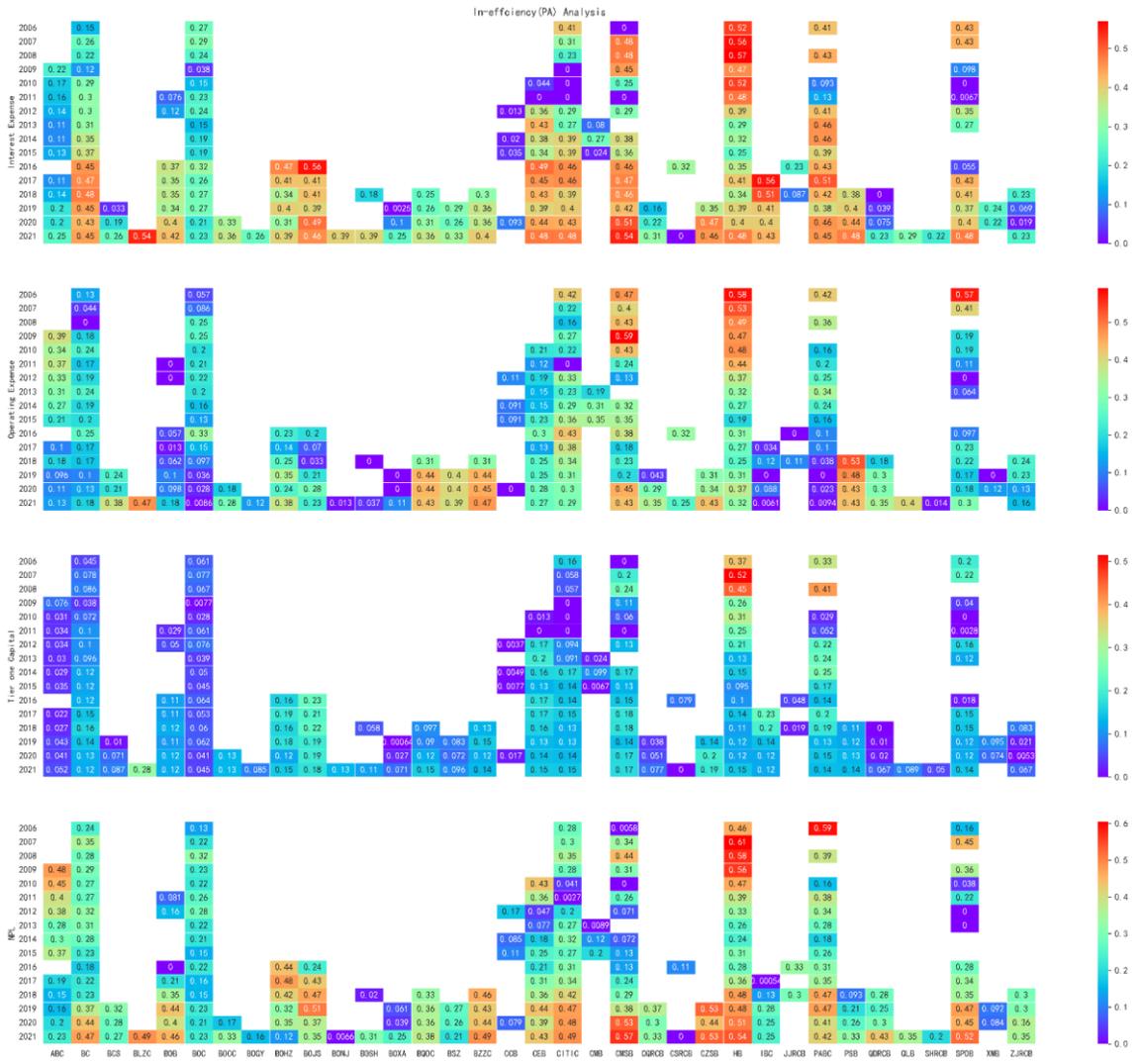

Figure 4: Probability Density of inefficiency (PA) of listed banks(2006-2021)

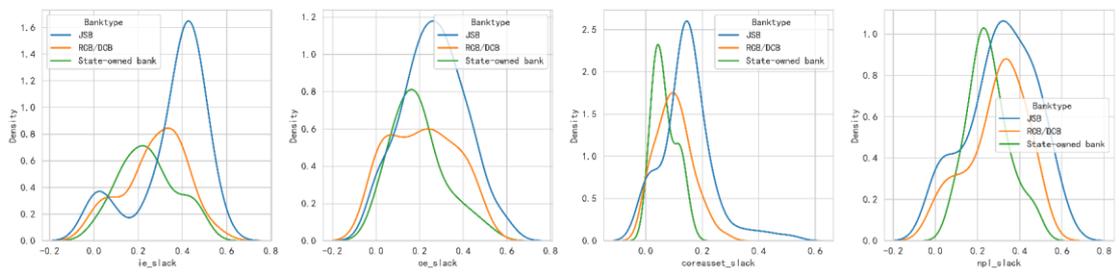

Figure 5: IA analysis chart of listed banks(2006-2021)

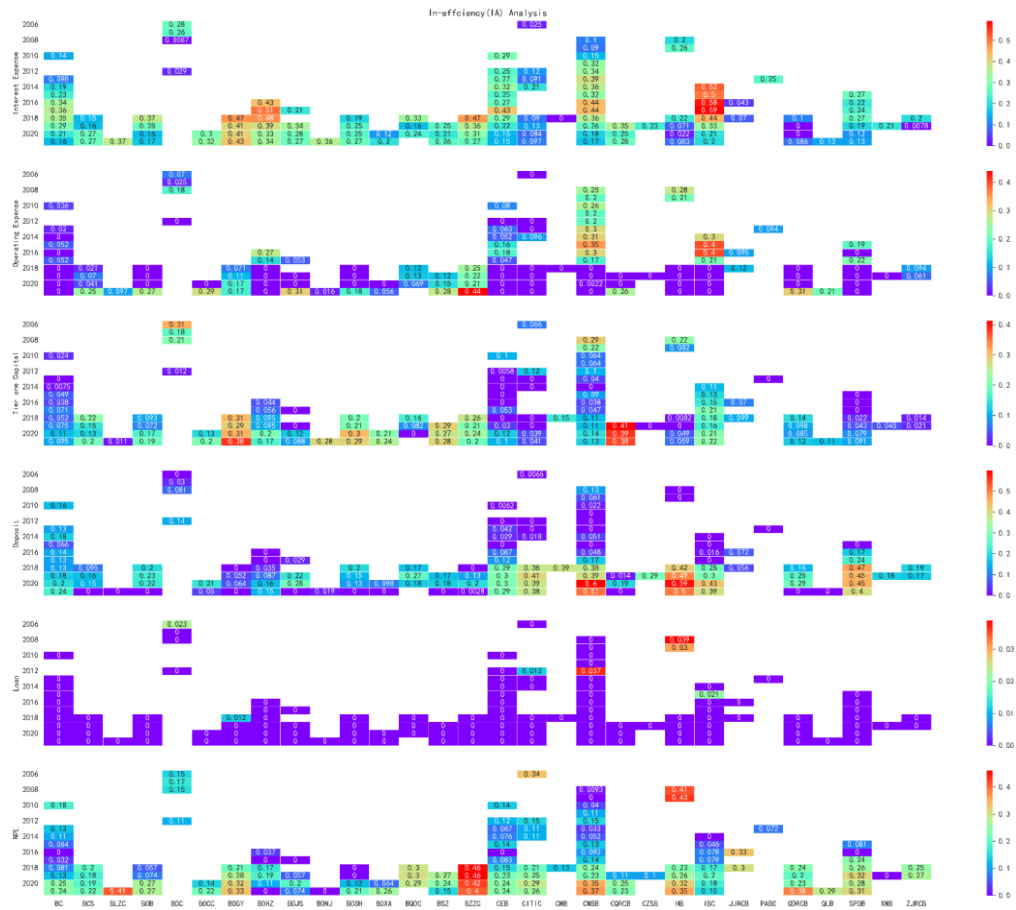

Figure 6: Probability Density of inefficiency (IA) of listed banks(2006-2021)

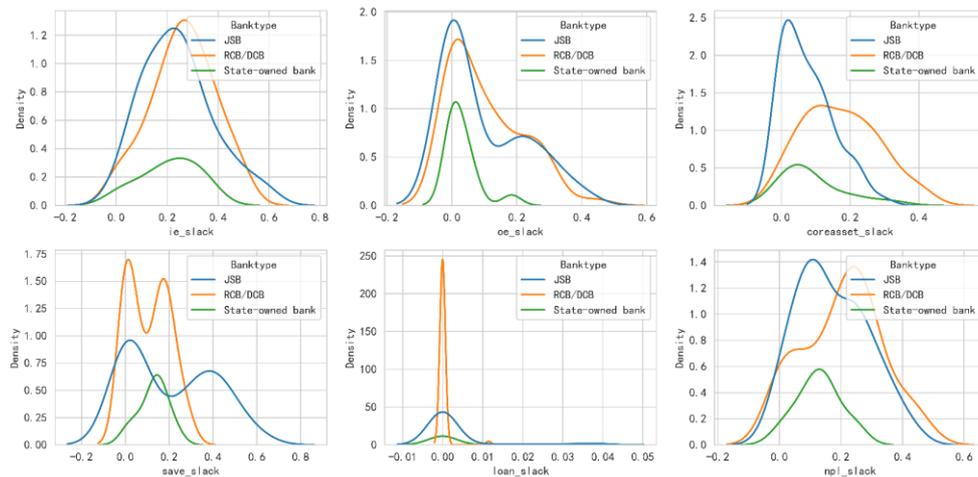

Figure 5 shows the inefficiency (IA) analysis diagram of listed banks. It can be seen from the figure that among joint-stock commercial banks, Minsheng Bank, Huaxia Bank, and regional banks as Zhengzhou Bank, Lanzhou Bank, Qingnong Commercial Bank and Guiyang Bank, the inefficiency of non-performing-loans is high; among joint-stock commercial banks, Minsheng Bank, Huaxia Bank, Shanghai Pudong Development Bank, Industrial Bank and China Citic Bank, the deposit inefficiency is high.

Regional commercial banks, such as Bank of Lanzhou, Bank of Nanjing, Bank of Hangzhou, Bank of Guiyang and Bank of Chongqing, are mainly inefficient in interest expenditure. From the IA inefficiency analysis, it can be seen that regional commercial banks have higher inefficiency of interest expenditure (which may be due to the use of higher interest expenditure in deposit absorption), while joint-stock commercial banks have higher inefficiency of deposit expenditure (that is, lower deposit absorption capacity).

### *4.3 Super efficiency decomposition - MI Analysis of commercial banks*

In order to study the efficiency changes of commercial banks from 2006 to 2021, we use the Malmquist index to demonstrate the competition and operation of banks, mainly capturing the changes in the efficiency of banks themselves and the changes in the overall production frontier of the banking sector. That is, the catch-up effect and the frontier shift effect. The change of the production frontier reflects the trend of the efficiency of the whole banking sector over time.

Figure 7 shows the MI (PA) index decomposition of listed banks. The upper row is catch-up effect, reflects the up and down of banks in industrial competition (of the same or similar scale) . A value greater than 1 indicates a rise in position, while a value less than 1 indicates a decline in position. The bottom row shows the change effect of production frontier, which reflects the improvement or decline of the overall operation level of banks of this scale. If the value is greater than 1, the overall efficiency of the banking sector is improved; if the value is less than 1, the overall efficiency is decreased.

Figure 7: Breakdown table of MI (PA) indicators of listed banks from 2006 to 2021

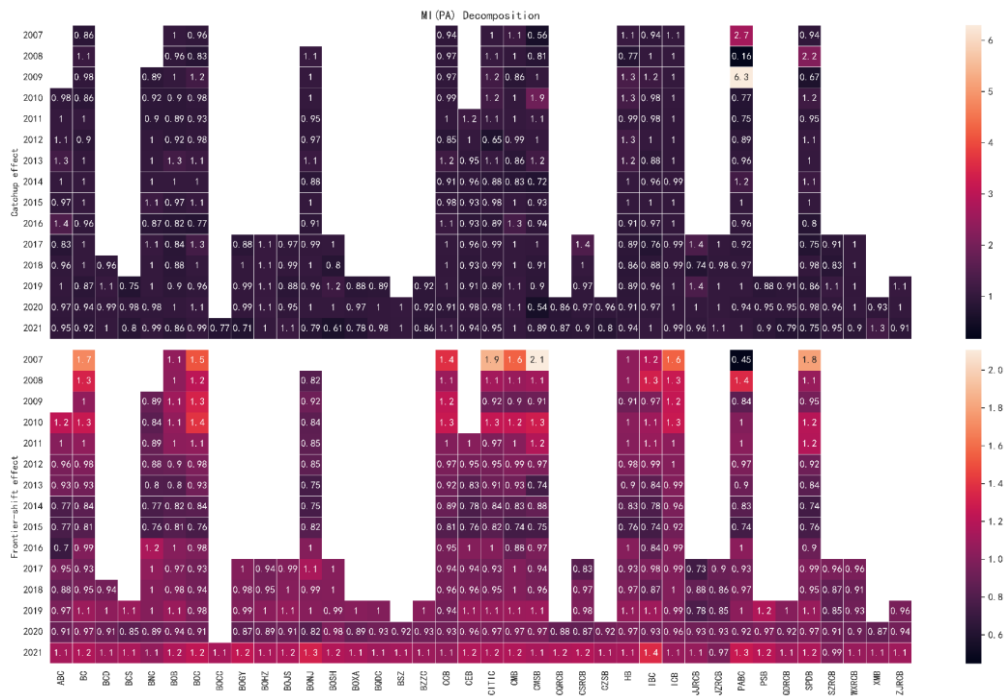

Figure 7 shows that the overall PA efficiency of the banking sector improved from 2007 to 2010, which may be attributed to the M2 growth rate and the effect of non-performing assets disposal. The overall efficiency decreased from 2011 to 2017, and developed steadily from 2018 to 2019. The impact of the COVID-19 outbreak in 2020 had a certain impact on the overall efficiency of the banking sector.

Figure 8 shows the MI (IA) index breakdown table of listed banks. It can be seen from the lower column of the figure that the banking industry expanded significantly from 2009 to 2010, while the overall PA and IA efficiency of the listed banking industry showed a downward trend from 2011 to 2015. The frontier efficiency declined the most in 2010, which may be related to the shrinkage of deposits and loans in the post-crisis era. The overall efficiency is relatively stable from 2016 to 2021.

Figure 9 shows the trend of MI index and frontier movement of the banking sector.

Figure 8: Breakdown table of MI (IA) indicators of listed banks from 2006 to 2021

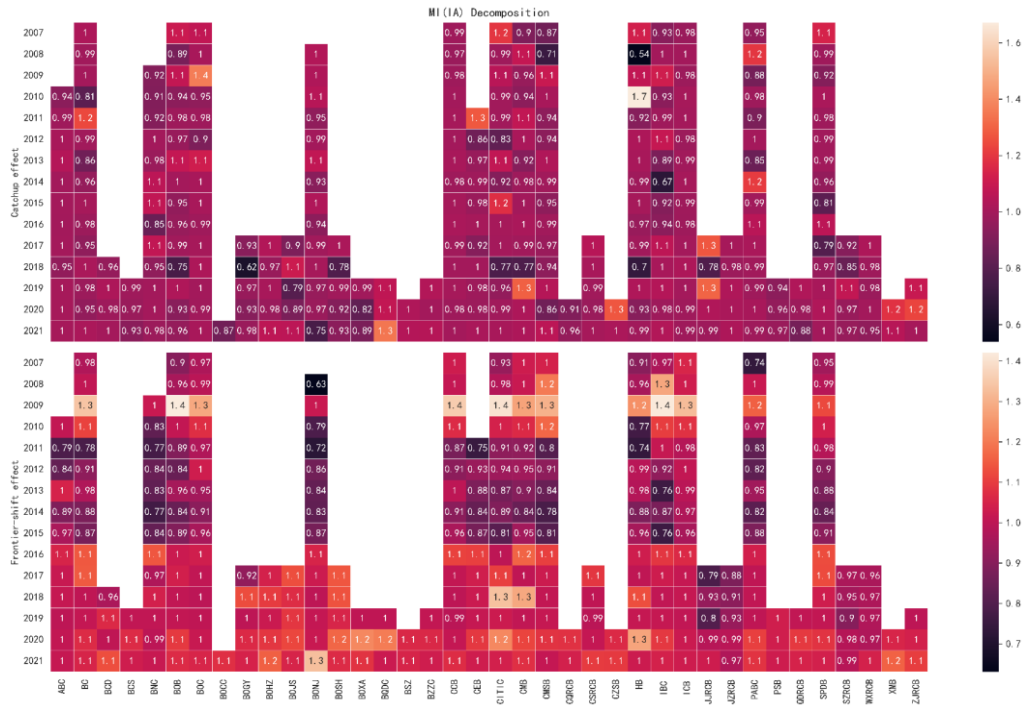

Figure 9: Trend of MI and Frontier–shift effect(2006-2021)

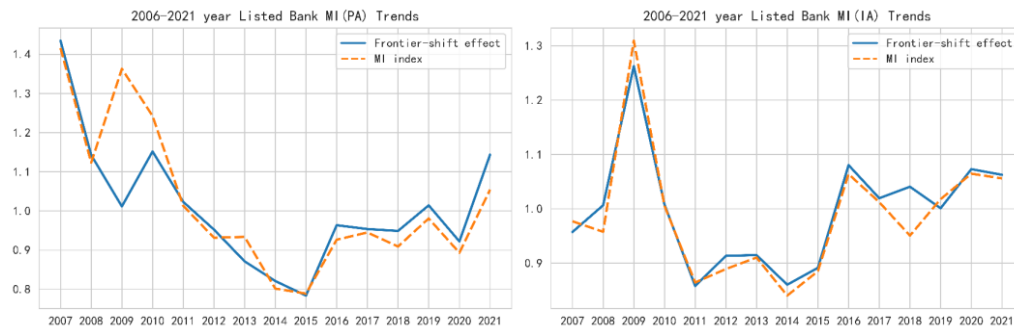

## 5.Analysis of the determinants of bank super efficiency

### *5.1 Model and variable selection*

The super efficiency model captures two types of factors: The catch-up effect and the frontier-shift effect represent the individual bank effort and the macro environment effect, which enables us to analyse the two effects separately. To avoid the bias caused by too many regressors and few samples, differencing panel model are used. Inspired by the analysis of Maudos, Guevara et al. (2004), Boyd and Nicolo (2005), we construct regression models as follows:

$$Catch_{it} - 1 = \sum_{k=1}^{k} \beta^k diff(BS_{i,t}^k) + \varepsilon_{it} \tag{5}$$

Catch variable is the catch-up effect, which measures the improvement or decline of the relative efficiency brought by the changes of individual banks effort. If the variable is larger than 1, it means the improvement of the relative efficiency; if it is smaller than 1, it means the decline of the efficiency. BS(bank-specific) variables are mainly bank specific variables.

FS variable is the production frontier effect, which measures the overall efficiency change of the banking sector, and ME(macroeconomic) variables are main influential factor, limited by the number of samples, we leave the part two of equation 4 to the future work.

Specific variables of banks include asset-liability ratio and loan-to-deposit ratio, which reflect the basic asset attributes of banks. The attributes reflecting risk control ability include loan concentration (the proportion of loan balance of the ten largest customers) and non-interest income (the proportion of non-interest income divided by total income) to measure income diversity; The attributes reflecting corporate governance include ownership concentration (sum of the squares of the proportion of the top ten shareholders) and foreign equity participation (the largest foreign shareholder divided by the proportion of the largest shareholder).

Table 5: Summary of Bank-specific Variables and Macroeconomic Variables

|  | N | Mean | SD | Min | p25 | Median | p75 | Max |
|---|---|---|---|---|---|---|---|---|
| dALR | 311 | -0.002 | 0.009 | -0.090 | -0.006 | -0.002 | 0.002 | 0.045 |
| dLDR | 311 | 0.032 | 0.056 | -0.177 | 0.001 | 0.032 | 0.057 | 0.292 |
| dtenclient | 311 | -0.027 | 0.392 | -0.831 | -0.155 | -0.065 | 0.039 | 6.020 |
| downhhi | 311 | 0.020 | 0.264 | -0.569 | -0.009 | 0.000 | 0.015 | 3.947 |
| dotheri | 311 | 0.075 | 1.002 | -16.472 | -0.070 | 0.097 | 0.278 | 1.973 |
| droe | 311 | 0.007 | 0.324 | -0.797 | -0.081 | -0.036 | 0.050 | 4.960 |
| dfown | 311 | -0.007 | 0.073 | -1.000 | 0.000 | 0.000 | 0.000 | 0.258 |
|  | N | Mean | SD | Min | p25 | Median | p75 | Max |
| spread | 16 | 0.021 | 0.009 | 0.005 | 0.015 | 0.018 | 0.027 | 0.039 |

| | | | | | | | |
|---|---|---|---|---|---|---|---|
| rf | 16 | 0.032 | 0.007 | 0.023 | 0.026 | 0.032 | 0.036 | 0.044 |
| cpi | 16 | 102.549 | 1.675 | 99.31 | 101.515 | 102.3 | 103.11 | 105.86 |
| m2r | 16 | 0.138 | 0.053 | 0.081 | 0.096 | 0.135 | 0.168 | 0.285 |
| lpr | 16 | 0.053 | 0.01 | 0.038 | 0.043 | 0.054 | 0.061 | 0.069 |
| rgdp | 16 | 0.083 | 0.028 | 0.023 | 0.068 | 0.079 | 0.096 | 0.142 |
| shiborv | 16 | 0.455 | 0.238 | 0.118 | 0.2 | 0.542 | 0.65 | 0.82 |
| aloancon | 16 | 0.053 | 0.008 | 0.04 | 0.045 | 0.055 | 0.061 | 0.062 |

### *5.3 Results of catch-up effect analysis*

Firstly, unit root test is conducted on all differencing variables, Arellano-Bond auto regressive effect is tested on explained variables. The reported statistics of PA catch-up effect AR[1,2] are z=[0.87,-0.40], and IA catch-up effect AR[1,2] is z=[-0.38,-1.06]. Auto regressions are rejected in both cases.

Models 1 to 3 and 4 to 6 in Table 6 are fixed-effect, random GLS and mean regression models respectively. From the preliminary model, the decline in LDR ratio and equity concentration have a significant impact on improving PA efficiency ($p<0.1$), and the decline in ownership concentration can also improve IA efficiency. From the regression, The IPO of commercial banks and the reform of ownership have promoted the overall efficiency.

As can be seen from Table 6, the differencing panel model has a high explanatory power on PA efficiency, but a lower explanatory power on IA efficiency, which reveals the IA efficiency are more related to the macro variables.

Further, the two groups of fixed and random effect models are tested by Hausmann test, and the results of Hausmann test are as follows: PA model: chi2(8) = 2.13; Prob > chi2 = 0.9767; IA model: chi2(8)=2.21; Prob > chi2 = 0.9740; Therefore, we conducted a panel regression with random effects for the three types of banks, and the results are shown in Table 7.

Models 1 to 3 respectively represent the PA catch-up regression results of state-owned banks, joint-stock banks and regional commercial banks, while models 4 to 6 represent the IA catch-up regression results. All results are reported as robust standard errors.

Table 6: Catch-up effect(pa) Panel Regression result (1)

| Models | (1) CatchPA | (2) CatchPA | (3) CatchPA | (4) CatchIA | (5) CatchIA | (6) CatchIA |
|---|---|---|---|---|---|---|
| dalr | 2.973 | 2.716 | 2.534 | .615 | 1.145 | 1.393 |
|  | (1.388) | (1.276) | (1.179) | (.676) | (1.15) | (1.345) |
| dcar | .217 | .228 | .247 | -.076 | -.04 | -.011 |
|  | (.911) | (.962) | (1.038) | (-.913) | (-.451) | (-.132) |
| dloantosave | -.231 | -.251 | -.237 | -.25 | -.224 | -.209 |
|  | (-1.154) | (-1.473) | (-1.513) | (-1.384) | (-1.28) | (-1.222) |
| dtenclient | -.01 | -.009 | -.007 | -.021 | -.021 | -.02 |
|  | (-.706) | (-.781) | (-.627) | (-1.554) | (-1.604) | (-1.451) |
| downhhi | **-.067**\*\*\* | **-.056**\*\*\* | **-.045**\*\* | **-.035**\* | **-.034**\* | **-.031**\* |
|  | (-3.243) | (-2.77) | (-2.143) | (-1.724) | (-1.773) | (-1.718) |
| dotheri | -.001 | -.001 | 0 | -.001 | -.002 | -.001 |
|  | (-.423) | (-.382) | (-.084) | (-.507) | (-.982) | (-.659) |
| droe | **.232**\*\*\* | **.24**\*\*\* | **.246**\*\*\* | -.008 | -.005 | .001 |
|  | (4.914) | (4.876) | (5.054) | (-.323) | (-.197) | (.025) |
| dfown | .079 | .071 | .077 | .025 | .02 | .007 |
|  | (.578) | (.551) | (.735) | (.535) | (.437) | (.146) |
| _cons | -.002 | -.002 | -.001 | .003 | .004 | .001 |
|  | (-.216) | (-.238) | (-.173) | (.431) | (.407) | (.088) |
| Observations | 309 | 309 | 309 | 309 | 309 | 309 |
| Pseudo $R^2$ | .z | .z | .z | .z | .z | .z |

*t-values are in parentheses*
*\*\*\* p<.01, \*\* p<.05, \* p<.1*

The results in Table 7 show that the PA catch-up effect of joint-stock banks significantly benefit from the decrease of ownership concentration and loan-to-deposit ratio, while the PA catch-up effect of regional commercial banks can significantly benefit from the decrease of customer concentration and the increase of reserves. For IA intermediation efficiency, large state-owned banks should increase the LDR ratio, while joint-stock banks should decrease the LDR ratio. Joint-stock banks benefit significantly from the decrease of ownership concentration, foreign capital introduction; RCB and CCB have benefited significantly from the decline in loan customer concentration.

Table 7: Catch-up effect(IA) Panel Regression result(2)

| Models | (1) CatchPA | (2) CatchPA | (3) CatchPA | (4) CatchIA | (5) CatchIA | (6) CatchIA |
|---|---|---|---|---|---|---|
| dalr | .259 | 5.309 | 2.511 | **2.152**** | -.099 | 2.334 |
|  | (.046) | (1.123) | (1.514) | (2.243) | (-.042) | (1.446) |
| dcar | -.563 | .398 | **.273**** | -.08 | **-.177*** | .198 |
|  | (-1.232) | (.748) | (2.081) | (-.489) | (-1.789) | (1.471) |
| dloantosave | .185 | **-.508*** | -.064 | **.421**** | **-.583**** | -.036 |
|  | (.736) | (-1.725) | (-.317) | (2.472) | (-2.334) | (-.132) |
| dtenclient | -.06 | -.016 | **-.012*** | .063 | .057 | **-.033***** |
|  | (-.786) | (-.164) | (-1.811) | (.592) | (.877) | (-6.71) |
| downhhi | -.06 | **-.074***** | .026 | .005 | **-.031**** | -.048 |
|  | (-.449) | (-3.014) | (.363) | (.074) | (-2.219) | (-.487) |
| dotheri | -.117 | -.002 | -.001 | -.024 | -.029 | -.002 |
|  | (-1.414) | (-.045) | (-.139) | (-1.067) | (-1.619) | (-.387) |
| droe | **.143*** | **.235***** | **.23*** | .095 | -.018 | .053 |
|  | (1.912) | (3.888) | (1.744) | (1.039) | (-1.046) | (.456) |
| dfown | -.235 | .189 | -.399 | -.125 | **.05**** | -.099 |
|  | (-1.092) | (1.18) | (-1.419) | (-.885) | (2.321) | (-.518) |
| _cons | .019 | .015 | -.022 | -.002 | .022 | -.01 |
|  | (1.26) | (.88) | (-1.634) | (-.375) | (1.252) | (-.704) |
| Observations | 75 | 118 | 116 | 75 | 118 | 116 |
| Pseudo $R^2$ | .z | .z | .z | .z | .z | .z |

*t-values are in parentheses \*\*\* p<.01, \*\* p<.05, \* p<.1*

## 5.4 Correlogram of frontier-shift effect and macroeconomic factors

Due to the small number of samples of the annual frontier efficiency movement factors obtained in this study, the conclusions drawn from the regression analysis may have huge bias, correlogram is given as a reference. Figure ten show that the top four variables with the highest absolute correlation coefficient with the PA frontier-shift (FSPA) are GDP growth rate, interest rate spread, CPI and LPR, the highest absolute correlation coefficient with the IA frontier-shift (FSIA) are Treasury bond interest rate, Shibor-interbank volatility, CPI and LPR.

Figure 10: Correlogram of Frontier-shift effect and Macroeconomic Variables

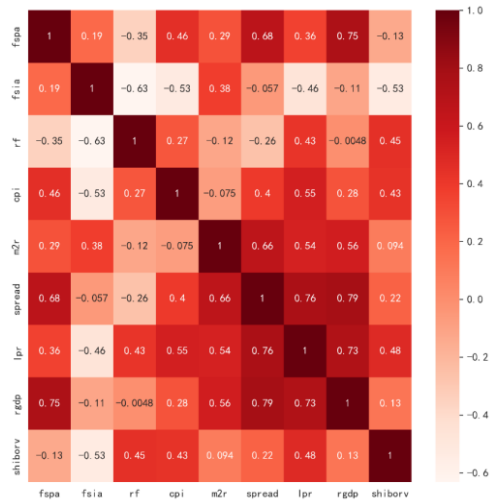

## 6. Conclusions and implications

We study the efficiency of China listed commercial banks which consist of 42 listed commercial banks in China from 2006 to 2021. By employed the Super-SBM-UND-VRS based DEA, accurate bank risk efficiency scores are obtained. The efficiencies obtained based on the production approach and the intermediary approach have different emphases, but both can comprehensively reflect the bank operating performance from the perspective of risk, and provide important measurement basis for high-quality development of China economy.

Some implications can be useful for stakeholders and regulators. For example, this paper reveals that the main sources of inefficiency of China joint-stock banks are interest expenses, operating expenses and non-performing assets, corresponding management should focus on loan approval and LDR management, lower down operating expenses. Second, regional commercial banks (RCB and CCB) should improve efficiency in interest expenses and reduce customer concentration. Third, large state-owned banks should make efforts to improve the efficiency of loans.

Secondly, this paper maybe the first academic paper to conduct differencing panel regression analysis on the MI index of super-efficiency decomposition, some enlightenment is obtained. For example, banks should improve their efficiency by controlling the LDR ratio and the concentration of loans. Regulators should monitor the risk of a decline in production frontier index, pay attention to the micro level of competition in the banking market, and encourage healthy competition. Our study also pointed out the IPO and reformation of ownership improve the overall banking efficiency. In addition, for the question of "competition to fragility" or "competition to stability", which has long been concerned in the field, this study may provide a verifiable index – frontier-shift effect index.

*References*

Andersen, P. and N. Petersen (1993). "A procedure for ranking efficient units in data envelopment analysis." Management science **39**(10): 1261-1264.

Banker, R. D., A. Charnes and W. W. Cooper (1984). "Some models for estimating technical and scale inefficiencies in data envelopment analysis." Management science **30**(9): 1078-1092.

Berger, A. N., I. Hasan and M. Zhou (2009). "Bank ownership and efficiency in China: What will happen in the world's largest nation?" Journal of Banking & Finance **33**(1): 113-130.

Boyd, J. H. and G. Nicolo (2005). "The Theory of Bank Risk-Taking and Competition Revisited." The Journal of Finance **60**(3).

Charnes, A., W. W. Cooper and E. Rhodes (1978). "Measuring the efficiency of decision making units." European journal of operational research **2**(6): 429-444.

Cooper, W. W., L. M. Seiford and K. Tone (2007). Data Envelopment Analysis, Springer Books.

Fang, H.-H., H.-S. Lee, S.-N. Hwang and C.-C. Chung (2013). "A slacks-based measure of super-efficiency in data envelopment analysis: An alternative approach." Omega **41**(4): 731-734.

Färe, R., S. Grosskopf, M. Norris and Z. Zhang (1994). "Productivity growth, technical progress, and efficiency change in industrialized countries." The American economic review: 66-83.

Fujii, H., S. Managi, R. Matousek and A. Rughoo (2018). "Bank efficiency, productivity, and convergence in EU countries: a weighted Russell directional distance model." The European Journal of Finance **24**(2): 135-156.

Gan, X. (2007). "SBM Analysis of the Efficiency of Chinese Commercial Banks - Controlling Macro and Ownership Factors. ." Financial Research (10A): 58-69.

Gao, J., X. Wang and X. Zhou (2021). "How does non interest income affect bank systemic risk: an empirical study based on mesomeric effect and panel threshold model." China Price(China) **No.391**(11): 67-70.

Grigorian, D. A. and V. Manole (2006). "Determinants of commercial bank performance in transition: an application of data envelopment analysis." Comparative Economic Studies **48**(3): 497-522.

Jiang, C., S. Yao and G. Feng (2013). "Bank ownership, privatization, and performance: Evidence from a transition country." Journal of Banking and Finance.

Jiang, X. and S. Li (2019). "Has non-interest income reduced the risk taking level of commercial banks—— Based on the perspective of scale heterogeneity." Journal of Regional Financial Research(China)(04): 11-16.

Li, N. (2021). "Non interest income, diversified income structure, and operational performance of commercial banks." Research on Financial Regulation(china) **118**(118): 76-93.

Li , Y. and Q. Wang (2018). "Removed red line: How deposit loan ratio constraints affect credit transmission of monetary policy." Finance & Economics(China)(6): 14.

Maudos, J., J. J. J. o. B. Guevara and Finance (2004). "Factors explaining the interest margin in the banking sectors of the European Union."

Sherman, H. D. and F. Gold (1985). "Bank branch operating efficiency: Evaluation with data envelopment analysis." Journal of banking & finance **9**(2): 297-315.

Tone, K. (2001). "A slacks-based measure of efficiency in data envelopment analysis." European journal of operational research **130**(3): 498-509.

Wang, K., W. Huang, J. Wu and Y.-N. Liu (2014). "Efficiency measures of the Chinese commercial banking system using an additive two-stage DEA." Omega **44**: 5-20.

Wang, X., X. Wang and Z. Bai (2020). "Research on the inverted U-shaped relationship between deposit loan ratio and profitability of commercial banks -- mesomeric effect model with non-performing loan ratio." Scientific research management(China) **41**(7): 9.

Yang, D. and A. Zhang (2007). "Efficiency Evaluation of Chinese Commercial Banks from 1996 to 2005: Empirical Analysis Based on Cost Efficiency and Profit Efficiency." Financial Research(China）(12A): 102-112.

Yao, S., G. Feng and C. Jiang (2004). "An Empirical Analysis of the Efficiency of China's Banking Industry." Economic Research(China) **8**(1): 111.

Zhang, D., J. Cai, D. G. Dickinson and A. M. Kutan (2016). "Non-performing loans, moral hazard and regulation of the Chinese commercial banking system." Journal of Banking and Finance **63**(feb.): 48-60.

Zhang, J. (2003). "DEA Method for Efficiency Research of Commercial Banks in China and Empirical Analysis of Efficiency from 1997 to 2001." Financial Research(China)(3): 11-25.

Zhao, D. (2021). "Non interest income and risk bearing level of Chinese commercial banks: empirical analysis based on dynamic panel SYS-GMM regression model." Technoeconomics & Management Research(China)(03): 68-72.

Zhao, S. and C. Shen (2016). The Impact of Developing Non Interest Business on Bank Returns and Risks -- An Empirical Study Based on 49 Commercial Banks in China. Economic Theory and Business Management(China). **2:** 83-97.

Zhu, N., B. Wang, Z. Yu and Y. Wu (2016). "Technical efficiency measurement incorporating risk preferences: An empirical analysis of Chinese commercial banks." Emerging Markets Finance and Trade **52**(3): 610-624.